\documentclass[noshowpacs,amsmath,
twocolumn,
superscriptaddress,
aps,prb]{revtex4-1}  % this article uses the Revtex 4-1 format
\usepackage{setspace}
\usepackage{amsmath}
\newcommand{\vect}[1]{\vec{#1}}    % Vectors use the arrow notation
\usepackage{verbatim}  % for commenting
\usepackage{amsfonts}
\usepackage{amssymb}
\usepackage[pdftex]{graphicx} % for pdf  figures
\usepackage{pdfpages}

\begin{document}

\title{Strong Interactions of Single Atoms and Photons near a Dielectric Boundary}

\author{D. J. Alton}
\thanks{These authors contributed equally to this work.}
\affiliation{Norman Bridge Laboratory of Physics MC 12-33,
California Institute of Technology, Pasadena, California 91125, USA}
\author{N. P. Stern}
\thanks{These authors contributed equally to this work.}
\affiliation{Norman Bridge Laboratory of Physics MC 12-33,
California Institute of Technology, Pasadena, California 91125, USA}
\author{Takao Aoki}
\affiliation{ Department of Physics, Kyoto University, Kyoto, Japan}

\author{H. Lee}
\affiliation{T. J. Watson Laboratory of Applied Physics MC 128-95, California Institute of Technology, Pasadena, California 91125, USA}
\author{E. Ostby}
\affiliation{T. J. Watson Laboratory of Applied Physics MC 128-95, California Institute of Technology, Pasadena, California 91125, USA}
\author{K. J. Vahala}
\affiliation{T. J. Watson Laboratory of Applied Physics MC 128-95, California Institute of Technology, Pasadena, California 91125, USA}
\author{H. J. Kimble}
\affiliation{Norman Bridge Laboratory of Physics MC 12-33,
California Institute of Technology, Pasadena, California 91125, USA}

%\date{\today}
\setstretch{.99}

%%%%%%%%%%%%%%%%%%%%%%%%%%%%%%%%%%%%%%%%%%%%%%%%%%%%%%%%%%%%%%%%%%%%%%%%%%%%
%%%%%%%%%%%%%%%%%%%%%%%%%%%%%%%%%%%%%%%%%%%%%%%%%%%%%%%%%%%%%%%%%%%%%%%%%%%%
%%%%%%%%%%%%%%%%%%%%%%%% Intro section %%%%%%%%%%%%%%%%%%%%%%%
%%%%%%%%%%%%%%%%%%%%%%%%%%%%%%%%%%%%%%%%%%%%%%%%%%%%%%%%%%%%%%%%%%%%%%%%%%%%

\begin{abstract}
Modern research in optical physics has achieved quantum control of strong interactions between a single atom and one photon within the setting of cavity quantum electrodynamics (cQED) \cite{Miller:2005}. However, to move beyond current proof-of-principle experiments involving one or two conventional optical cavities to more complex scalable systems that employ $N\gg 1$ microscopic resonators \cite{Kimble:2008} requires the localization of individual atoms on distance scales $\lesssim 100$nm from a resonator's surface. In this regime an atom can be strongly coupled to a single intracavity photon \cite{Aoki:2006} while at the same time experiencing significant radiative interactions with the dielectric boundaries of the resonator \cite{HarocheKleppner:1989}. Here, we report an initial step into this new regime of cQED by way of real-time detection and high-bandwidth feedback to select and monitor single Cesium atoms localized $\sim 100$ nm from the surface of a micro-toroidal optical resonator. We employ strong radiative interactions of atom and cavity field to probe atomic motion through the evanescent field of the resonator. Direct temporal and spectral measurements reveal both the significant role of Casimir-Polder attraction \cite{Casimir-Polder:1948} and the manifestly quantum nature of the atom-cavity dynamics. Our work sets the stage for trapping atoms near micro- and nano-scopic optical resonators for applications in quantum information science, including the creation of scalable quantum networks composed of many atom-cavity systems that coherently interact via coherent exchanges of single photons \cite{Kimble:2008}.
\end{abstract}
\maketitle

The proximity of dielectric boundaries fundamentally alters atomic radiative processes as compared to quantum electrodynamics in free space. For example, free-space Lamb shifts and Einstein-$A$ coefficients (i.e., level positions and decay rates) are modified for atom-surface distances comparable to the relevant transition wavelengths, as considered in the pioneering analyses of Casimir and Polder \cite{Casimir-Polder:1948} and of Purcell \cite{Purcell:1946} in the late 1940s. Seminal experiments in the 1970s investigated radiative decay for organic dye molecules near a metal mirror \cite{Drexhage:1974} and were followed in the 1980s by landmark observations of the inhibition of spontaneous emission for a trapped electron \cite{Gabrielse:1985} and an atom in a waveguide \cite{Hulet:1985}. The ensuing years have witnessed the development of cavity quantum electrodynamics (cQED) in this \emph{perturbative} regime of boundary-modified linewidths and level shifts \cite{HarocheKleppner:1989, Sukenik:1993, Berman:1994}, with applications ranging from measurements of fundamental constants \cite{Odom:2006} to the development of novel semiconductor devices \cite{devices}.

With increased interaction strength, a \emph{non-perturbative} regime of cQED becomes possible and is characterized not by irreversible decay but rather by the cyclic, reversible exchange of excitation between atom and photon \cite{JaynesCummings:1963}. The experimental quest for strong atom-photon coupling had its initial success in 1985 in the microwave regime with the realization of the micromaser \cite{Meschede:1985}, with strong coupling in the optical domain achieved some years later \cite{Thompson:1992}. By now the coherent control of atomic radiative dynamics has become possible on a photon-by-photon basis\cite{Miller:2005,HarocheBook}. Strong coupling has also been demonstrated for a wide class of physical systems \cite{Vahala:2004} beyond single atoms, including quantum dots coupled to micropillars and photonic bandgap cavities \cite{Khitrova:2006} and Cooper-pairs interacting with superconducting resonators \cite{Schoelkopf:2008, Hofheinz:2008}. This non-perturbative regime of cQED with strong light-matter interactions mediated by single photons has led to new scientific capabilities, ranging from a laser that operates with one-and-the-same atom \cite{McKeever:2003} to the deterministic generation of entangled photon pairs \cite{Rempe:2009} to a two-qubit superconducting quantum processor \cite{DiCarlo:2009}.

To a large degree, advances in the perturbative and non-perturbative regimes of cQED have been made independently. For example, for one atom localized near the center of a Fabry-Perot cavity with volume $(l)^3 \sim (10$ $\mu$m$)^3$, the coherent coupling $g$ to an optical resonance can be large compared to radiative decay characterized by the Einstein-$A$ coefficient and cavity loss rate $\kappa$, namely  $g\gg (\gamma_0,\kappa)$ where $\gamma_0 = A/2$, placing the system in the regime of strong, non-perturbative atom-photon coupling \cite{Miller:2005}. Nevertheless, corrections to the atomic Lamb shift and Einstein-$A$ coefficient arising from surface interactions with the cavity boundaries remain small (e.g., $\delta A/A \sim 10^{-5}$). However, many applications in Quantum Information Science could benefit from strong atom-photon interactions with micro- and nano-scopic optical resonators \cite{Folman:2002, Armani:2003, Spillane:2005, Eichenfeld:2009}. Atomic localization on a sub-wavelength scale near a resonator's surface is then required, with aspects of both perturbative and non-perturbative cQED necessarily coming into play.

\begin{figure}
\includegraphics[width=1.0\columnwidth]{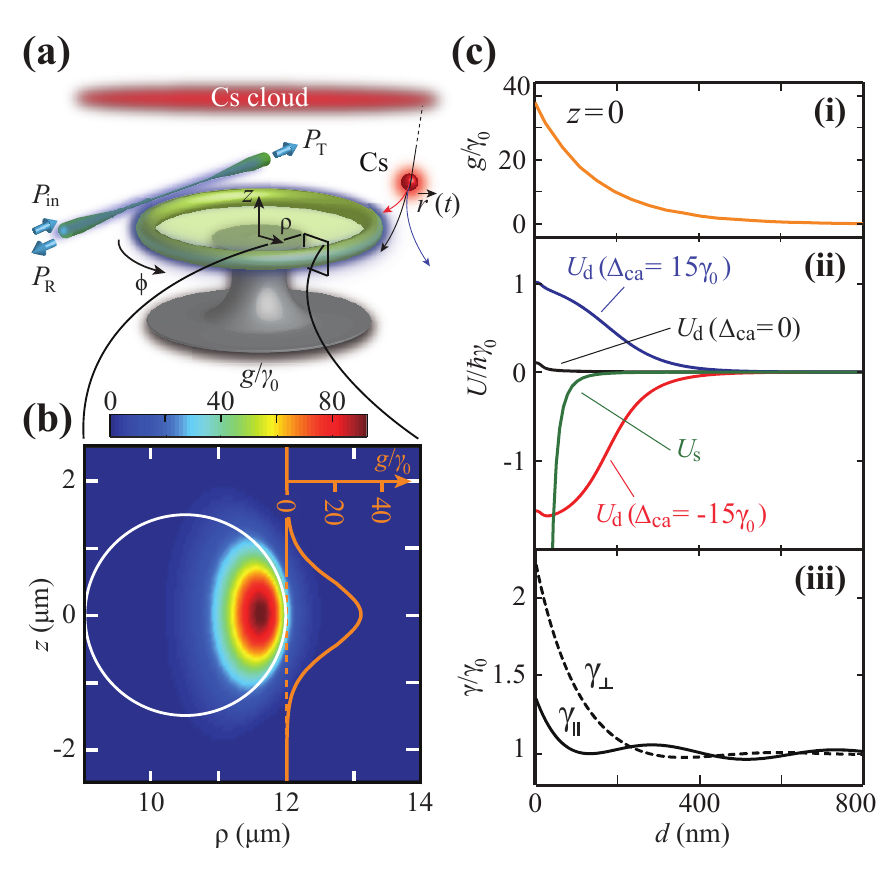}\vspace{-0.10in}\caption{\label{fig1}
\textbf{Radiative interactions and optical potentials for an atom near the surface of a toroidal resonator. a,} Simple overview of the experiment showing a cloud of cold Cesium atoms that is released with then a few atoms falling near a microtoroidal resonator. Light in a tapered optical fiber excites the resonator with input power $P_{\rm in}$ at frequency $\omega_{\rm p}$, leading to transmitted and reflected outputs $P_{\rm T},P_{\rm R}$. \textbf{b,} Cross section of the microtoroid at $\phi=0$ showing the coherent coupling coefficient $|g\left(\vect{r}\right)=g(\rho,z,\phi)|$ for a TE polarized whispering-gallery mode. The microtoroid has principal and minor diameters $\left(D_{\rm p}, D_{\rm m}\right) = \left(24, 3\right)$ $\mu$m, respectively. \textbf{c,} (i) Coherent coupling $|g(d,z,\phi)|$ for the external evanescent field as a function of distance $d = \rho - D_{\rm p}/2$ from the toroid's surface for $\left(z, \phi\right) = \left(0, 0\right)$. (ii) The effective dipole potentials $U_{\rm d}$ for resonant $\omega_{\rm p} = \omega^{(0)}_{\rm a}$, red $\omega_{\rm p} < \omega^{(0)}_{\rm a}$ and blue $\omega_{\rm p} > \omega^{(0)}_{\rm a}$ free-space detunings of the probe $P_{\rm in}$. The Casimir-Polder surface potential $U_{\rm s}$ for the ground state of atomic Cs is also shown. (iii) The atomic decay rate $\gamma(d)$ as a function of distance $d$ from the toroid's surface for TE ($\gamma_\|$) and TM ($\gamma_\bot$) modes. All rates in this figure are scaled to the decay rate in free space for the amplitude of the Cs $6P_{3/2} \rightarrow 6S_{1/2}$ transition, $\gamma_0/2\pi = 2.6$ MHz. The approximate distance scale probed in our experiment is $0 < d < 300$ nm.
}\end{figure}

In this manuscript we investigate such a regime for single Cesium atoms radiatively coupled to a high-$Q$ microtoroidal cavity\cite{Aoki:2006,Armani:2003} and localized near the resonator's dielectric surface. As illustrated in Fig.~\ref{fig1}a, cold Cesium atoms are released from an optical dipole-force trap and randomly fall past the microtoroid. A real-time detection scheme based upon strong radiative interactions between one atom and the evanescent field of the cavity selects atomic trajectories localized within $d\lesssim300$ nm from the resonator's surface, with a large fraction of atoms passing below $100$ nm and crashing into the surface. On this scale, the atom's coherent interaction with the cavity field is characterized by strong, non-perturbative coupling [Fig. \ref{fig1}b, \ref{fig1}c(i)], which we demonstrate by direct measurements of so-called ``vacuum-Rabi'' spectra for light transmitted and reflected by the atom-cavity system, as well as by observations of photon antibunching for the transmitted light. On the other hand, the atom's motion and level structure are significantly influenced by the (perturbative) Casimir-Polder potential from the surface's proximity [Fig. \ref{fig1}c(ii)], which we infer from measurements of the time dependence of the cavity transmission during an atomic transit event, as well as from modifications of the spectra recorded for the transmitted and reflected fields. These observations are in reasonable agreement with a theoretical model that we have implemented by Monte-Carlo simulation and which gives insight into the underlying atomic dynamics, as detailed in Ref.~\onlinecite{SimPaper}.

%fig1

For the identification of atoms near the surface of the microtoroid in the regime shown in Fig. \ref{fig1}c, we rely on the strong interaction of atom and cavity field to modify the light transmitted by the cavity. Specifically, because the atom-cavity coupling coefficient $g\left(\vect{r}\left(t\right)\right)$ depends upon the atomic trajectory $\vect{r}(t)$, we can select single atoms localized in the cavity mode by demanding a minimum criterion for the change in cavity transmission due to the atomic trajectory. Our scheme for single-atom detection is similar to that used in previous work\cite{Aoki:2006,Dayan:2008,Aoki:2009}, but with significant modifications. Namely, by implementing fast digital logic, we achieve reliable real-time identification of atomic transit events in times as short as $250$ ns from the photoelectric counts from the transmitted power $P_{\rm T}(t)$. Given the identification of a localized atom, the control logic switches the power $P_{\rm in}$ and frequency $\omega_{\rm p}$ of the probe input within $\simeq 100$ ns and records subsequent photoelectric counts for the transmitted $P_{\rm T}(t)$ and reflected $P_{\rm R}(t)$ outputs from the cavity. These records of photoelectric counts form the basis for our analysis that follows, with further details presented in the Appendix and Supplementary Information (SI).

%%%%%%%%%%%%%%%%%%%%%%%%%%%%%%%%%%%%%%%%%%%%%%%%%%%%%%%%%%%%%%%%%%%%%%%%%%%%
%%%%%%%%%%%%%%%%%%%%%%%% Temporal Dynamics section %%%%%%%%%%%%%%%%%%%%%%%
%%%%%%%%%%%%%%%%%%%%%%%%%%%%%%%%%%%%%%%%%%%%%%%%%%%%%%%%%%%%%%%%%%%%%%%%%%%%

\begin{figure}
\includegraphics[width=1.0\columnwidth]{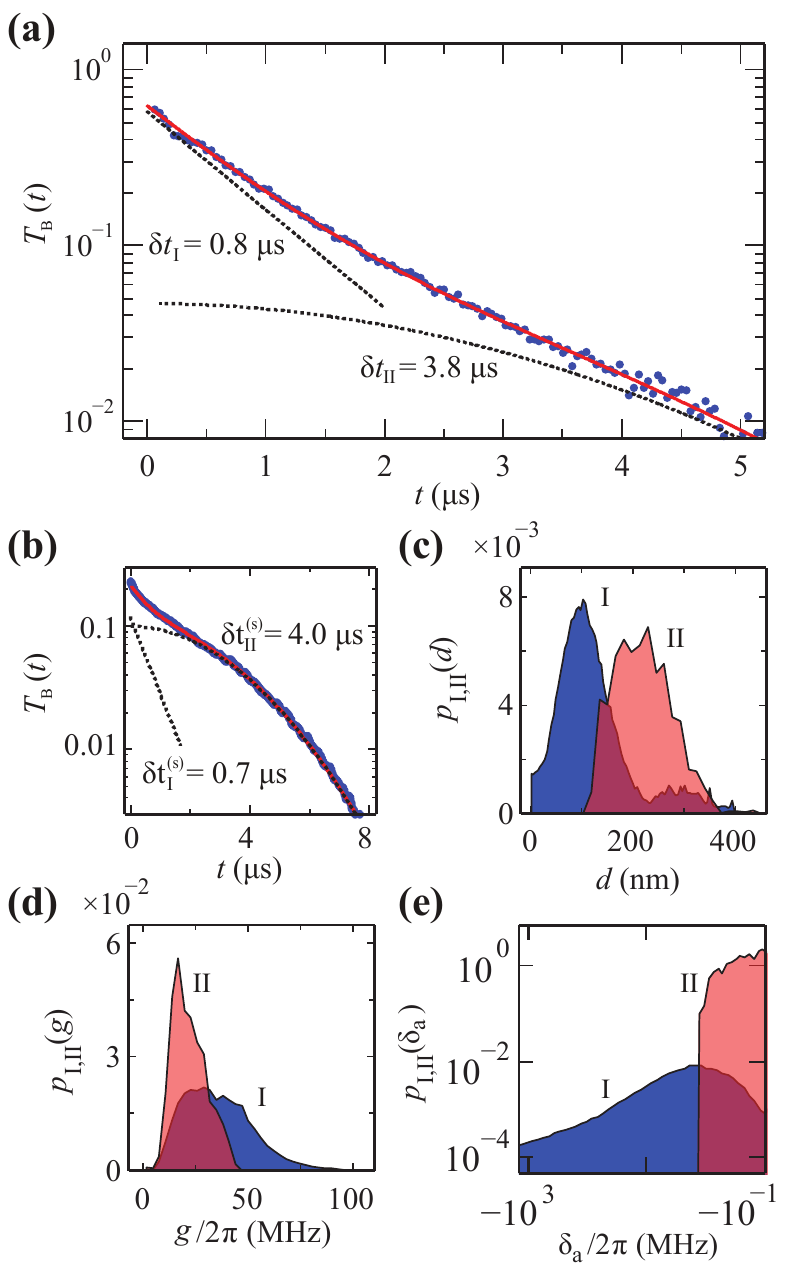}\vspace{-0.15in}\caption{\label{fig2}
\textbf{Observation (a) and simulation (b-e) of atomic transits within the evanescent field of the micro-toroidal resonator for $\Delta_{ca}=\Delta_{pa}=0$. a,} Observed cavity transmission $T_B(t)$ versus time $t$ following a triggering event at $t=0$, with approximately $5\times 10^4$ triggered transits included. The data are fit to the sum of an exponential (I) and a Gaussian (II) (red curve), with time constants $\delta t_{\rm I} = 0.78\pm0.02$ $\mu$s and $\delta t_{\rm II} = 3.75 \pm 0.09$ $\mu$s, with each component shown by the dotted lines. \textbf{b,} Simulation result for the cavity transmission $T^{(s)}_B(t)$ versus time $t$ (points) from an ensemble of triggered trajectories. The red curve is a fit to the sum of an exponential and Gaussian with time constants $\delta t^{(s)}_{\rm I} = 0.69$ $\mu$s, $\delta t^{(s)}_{\rm II} = 4.0$ $\mu$s while the dotted lines represent the individual fit components. \textbf{c-e} Probability densities $p_i(d),p_i(g),p_i(\delta_{\rm a})$ for the distance $d$, coupling $g$, and transition frequency shift $\delta_{\rm a} =  \omega_{\rm a}(d) - \omega^{(0)}_{\rm a}$ from the same simulation set as for (b). $\{d,g,\delta_{\rm a}\}$ are averaged over the first $500$ ns following the trigger. For these results, the trajectories are divided into two classes, $i=\{{\rm I},{\rm II}\}$ corresponding to the two time constants $\delta t^{(s)}_{\rm I}$ (blue shaded curve) and $\delta t^{(s)}_{\rm II}$ (red shaded curve) in (b).
}\end{figure}

To address experimentally the question of the distance scale for the recorded atom transit events, we first examine the time dependence of the cavity transmission $T(t)$ immediately following a trigger heralding the arrival of an atom into the cavity mode. Figure \ref{fig2}a shows $T(t) = P_{\rm T}(t)/P_{\rm in}$ for the case of resonant excitation, namely $\Delta_{\rm pa} = \omega_{\rm p} - \omega^{(0)}_{\rm a} = 0$ and $\Delta_{\rm ca} = \omega_{\rm c} - \omega^{(0)}_{\rm a} = 0$, where $\omega^{(0)}_{\rm a}$ is the free-space atomic frequency for the $6S_{1/2}, F=4 \rightarrow 6P_{3/2}, F=5$ transition in atomic Cs and $\omega_{\rm c}$ is the resonant frequency of the toroidal cavity. Two characteristic decay times are evident, with the background subtracted transmission $T_B(t) \equiv T(t) - B$ fitted well by the sum of an exponential ($\propto e^{-t/\delta t_I}$) with $\delta t_{\rm I} = 0.78 \pm0.02$ $\mu$s and a Gaussian ($\propto e^{-(t/\delta t_{II})^2}$) with $\delta t_{\rm II} = 3.75 \pm 0.09$ $\mu$s. Here, the background level $B \equiv T(t \gg \delta t_{I,II})$ is determined from the cavity transmission for times long compared to the duration of the transit event.

The time constants $\delta t_{\rm I}, \delta t_{\rm II}$ can be associated with distance scales $d_{\rm I}, d_{\rm II}$ by way of the average velocity $\bar{v}$ with which atoms arrive at the toroid's mode following release from the optical trap, namely $\bar{v} \sim 0.17$ m/s, leading to $d_{\rm I} \simeq 130$ nm and $d_{\rm II} \simeq 640$ nm. For comparison, the scale length for $g(d)$ in the radial direction is $\lambdabar =1/k_0 = 136$ nm (Fig. \ref{fig1}c(i)), while in the vertical direction, the variation of $g(z)$ is approximately Gaussian  ($\propto e^{-(z/w_0)^2}$) with waist $w_0 \simeq 590$ nm (Fig. \ref{fig1}b). The comparisons $d_{\rm I} \sim \lambdabar$ and $d_{\rm II} \sim w_0$ suggest that the short-lived component $\delta t_{\rm I}$ in Fig. \ref{fig2}a arises from atomic trajectories that are deflected from their otherwise vertical fall to largely radial paths of length $\lambdabar$ that terminate at the dielectric surface. Similarly, the longer-lived component $\delta t_{\rm II}$ is associated with trajectories that pass along $z$ without significant radial motion toward the surface of the toroid.

%fig2

Of course an atom near the surface will not move with constant velocity but will be accelerated by interactions with surface potentials and the cavity field itself. To reach a quantitative understanding of the external, center-of-mass motion and the internal, atomic dipole-cavity field coupling, we have implemented a numerical simulation that incorporates both perturbative and non-perturbative aspects of the radiative interaction of the atom and micro-toroid. Our Monte-Carlo simulation draws random initial trajectories for atoms falling from a thermal cloud and implements a stochastic process for photoelectric detection to emulate our real-time detection technique. The model includes Casimir-Polder and dipole forces from the potentials $U_{\rm s}(\vect{r}), U_{\rm d}(\vect{r})$ shown in Fig. \ref{fig1}c(ii), atomic level shifts (and hence detunings) from $U_{\rm s}(\vect{r})$, and boundary-modified decay $\gamma_{\|}(d)$ shown in Fig. \ref{fig1}c(iii). The non-perturbative interaction of atom and cavity field is based upon the analytic results in Ref.~\cite{ Aoki:2006, Dayan:2008}. Details of the simulation appear in Ref.~\onlinecite{SimPaper} and the SI.

Results from this analysis are presented in Fig. \ref{fig2}(b-e). In agreement with the observations in Fig. \ref{fig2}a, $T_B(t)$ from the simulation in Fig. \ref{fig2}b exhibits two time scales and is fit well by the sum of an exponential with time constant $\delta t^{(s)}_{\rm I} = 0.69$ $\mu$s and a Gaussian with $1/e$ width $\delta t^{(s)}_{\rm II} = 4.0$ $\mu$s. Atomic trajectories associated with the shorter time $\delta t^{(s)}_{\rm I}$ have distances peaked around $d_{\rm I} \sim 100$ nm and terminate with crashes into the surface of the toroid [$p_{\rm I}(d)$ in Fig. \ref{fig2}c], but exhibit large coupling $g_{\rm I}/2\pi \sim 40$ MHz [$p_{\rm I}(g)$ in Fig. \ref{fig2}d] and large surface-induced shifts of the atomic transition frequency $\delta_{\rm a,I}/2\pi \gtrsim 10$ MHz for $d \lesssim 60$ nm [$p_{\rm I}(\delta_{\rm a})$ in Fig. \ref{fig2}e]. By contrast, the complementary set of trajectories for the longer time $\delta t^{(s)}_{\rm II}$ pass roughly vertically through the cavity mode along $z$ (hence their Gaussian time dependence). Relative to the $\delta t^{(s)}_{\rm I}$ set, these trajectories exhibit larger distances $d_{\rm II} \sim 250$ nm, smaller coupling $g_{\rm I}/2\pi \sim 20$ MHz, and surface-induced shifts of the atomic transition frequency $\delta_{\rm a,I} \lesssim \gamma_0$ [$p_{\rm II}(d),p_{\rm II}(g),p_{\rm II}(\delta_{\rm a})$ in Figs. \ref{fig2}(c-e), respectively].

\begin{figure*}
\begin{center}
\includegraphics[width=16.5cm]{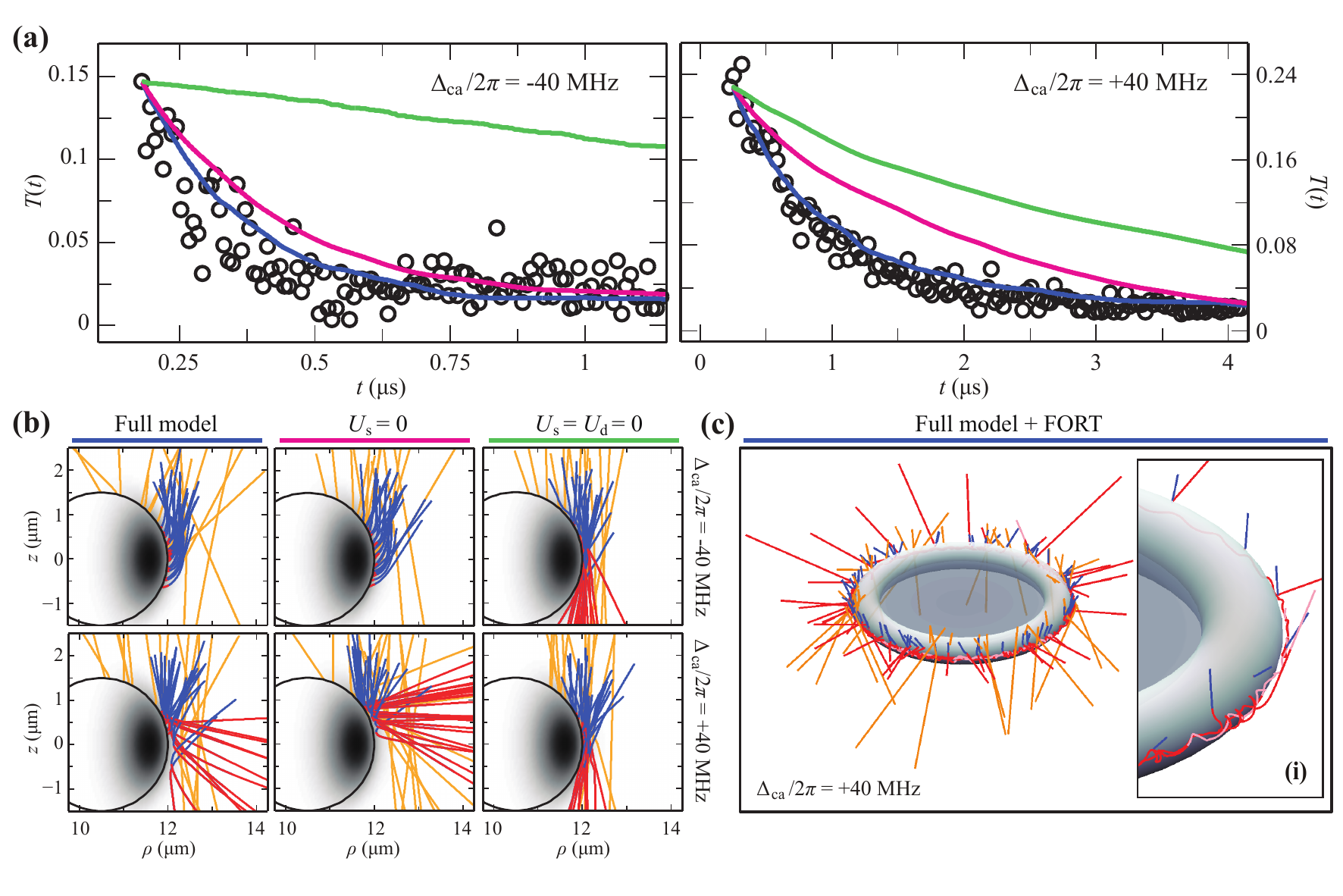}\end{center}\vspace{-0.25in}\caption{\label{fig3}
\textbf{Dynamics and trajectories for strongly coupled atoms moving in surface and dipole potentials $\{U_{\rm s},U_{\rm d}\}$. a,} Transmission $T(t)$ for $\Delta_{\rm ca}/2\pi = -40$ MHz (left) and $+40$ MHz (right) measured after an atom trigger at $t=0$.  In each panel, the circles are data for $2\times 10^3$ trigger events; the lines are simulations of $T(t)$ for the full model (blue), for $U_{\rm s}=0$ (magenta), and for $U_{\rm s} =U_{\rm d}=0$ (green). \textbf{b,} Representative atomic trajectories projected onto the $\rho-z$ plane for simulations in panel (a), with the TE mode intensity plotted on a gray scale. The upper panels are for $\Delta_{\rm ca}/2\pi = -40$ MHz while the lower panels are for $\Delta_{\rm ca}/2\pi = +40$ MHz. The color bars at the top of the panels match the colors of the curves in (a). For each panel, orange lines are untriggered trajectories, while triggered trajectories are represented by blue lines which turn red after a trigger at $t=0$.  \textbf{c,} Simulations showing trajectories from a full 3D simulation with $U_{\rm s}$, $U_{\rm d}$, as well as a two-color dipole potential (FORT) triggered ``on" by atom detection at $t = 0$. $\Delta_{\rm ca}/2\pi = +40$ MHz in correspondence to (a), (b). Blue lines represent falling atoms with the FORT beams ``off" $(t < 0)$, while red lines are trajectories after the FORT is triggered ``on" and an atom begins to orbit the toroid. To illustrate the timescale, the trajectories are colored pink for $t > 50$ $\mu$s.
}\end{figure*}

In Fig. \ref{fig3} we investigate temporal dynamics for the cavity transmission $T(t)$ but now with non-zero detuning between the atom and probe field, $\Delta_{\rm ca}/2\pi = \Delta_{\rm pa}/2\pi = \pm 40$ MHz (Fig. 3a).  Since $\omega_{\rm p} \neq \omega_{\rm a}$, dipole forces from coherent excitation of the intracavity field should induce an asymmetry for $T(t)$ for red and blue detuning, with faster decay for red detuning ($\omega_{\rm p} < \omega_{\rm a}$) due to the combined effect of the attractive potentials $U_{\rm s}(\vect{r})$ and $U_{\rm d}(\vect{r})$ shown in Fig. \ref{fig1}c(ii). The data are fit well by decaying exponentials ($\propto e^{-t/\delta t_i}$) added to Gaussian components in analogy with the earlier analysis, with exponential time constants $\delta t_{\rm red} = 0.11 \pm0.01$ and $\delta t_{\rm blue} = 0.53\pm0.03$ $\mu$s.  Also plotted are simulations of the experiment for freely falling atoms with $U_{\rm s} = U_{\rm d} = 0$, atoms under the influence of only dipole forces, $U_{\rm s} = 0$, and a full simulation including both the dipole force and CP forces. For both red and blue detunings, the timescales from the simulation absent $U_{\rm s},U_{\rm d}$ are substantially longer than observed in experiment, while the full simulation with $U_{\rm s}, U_{\rm d}$ leads to time constants $\delta t^{(s)}_{\rm red} = 0.19\pm0.02$ $\mu$s and $\delta t^{(s)}_{\rm blue} = 0.59\pm0.06$ $\mu$s.  These simulated time constants are qualitatively similar to observations, with quantitative differences attributed to simplifications inherent in the simulation model (see SI).  In contrast with the case $\omega_{\rm p}=\omega_{\rm a}$ in Fig.~\ref{fig2}, the Gaussian component of these temporal decays is minimal because of the difference in scale lengths of the CP potential $U_{\rm s}$ and effective dipole potential $U_{\rm d}(\Delta_{ca} =\pm 15\gamma_0)$, which become comparable to $\gamma_0$ for distances $d\lesssim \{65, 200\}$ nm, respectively (Fig. \ref{fig1}c(ii)).  Long range dipole forces, which are largely absent for Fig.~\ref{fig2}, dominate the trajectory dynamics of Fig.~\ref{fig3} and consequently vertically falling long-lived Gaussian trajectories do not significantly contribute.

To illustrate the underlying atomic motion, Fig.~\ref{fig3}b displays atomic trajectories projected onto the $\rho-z$ plane for each simulation in Fig. \ref{fig3}a.  Each panel displays a representative sample of untriggered and triggered trajectories. For red detuning, introducing dipole forces and CP forces leads to every triggered atom crashing into the toroid surface, explaining the short decay $\delta t_{\rm red}$.  The blue detuned case is more complicated, with both attractive CP forces and the repulsive dipole force reducing the time the atom is in the mode; CP forces pull nearby atoms into the surface while the dipole force repels other atoms out of the mode.

%%%%%%%%%%%%%%%%%%%%%%%%%%%%%%%%%%%%%%%%%%%%%%%%%%%%%%%%%%%%%%%%%%%%%%%%%%%%

As shown in Fig.~\ref{fig3}c and discussed in the Appendix, we have augmented our numerical simulation to include a dipole force optical trap $U_{\rm t}$ (FORT) formed by the toroid's evanescent field~\cite{Mabuchi:1994, Vernooy:1997, Rosenblit:2006} in addition to the potentials $U_{\rm s}$, $U_{\rm d}$.  The trapping potential $U_{\rm t}$ is triggered on by the same criteria as for Figs.~\ref{fig3}a, b, with then a significant fraction of triggered atoms bound in orbit around the toroid for durations surpassing 50 $\mu$s.

%%%%%%%%%%%%%%%%%%%%%%%% Spectra Section %%%%%%%%%%%%%%%%%%%%%%%
%%%%%%%%%%%%%%%%%%%%%%%%%%%%%%%%%%%%%%%%%%%%%%%%%%%%%%%%%%%%%%%%%%%%%%%%%%%%

The measurements in Figs. \ref{fig2}, \ref{fig3} rely upon strong interactions of one atom and photon for initial atomic localization within the cavity mode and for measurements of the subsequent motion by way of $T(t)$. To establish directly the non-perturbative coupling of atom and cavity field, we next turn to measurements of transmission $T(\omega_{\rm p}) = P_{\rm T}(\omega_{\rm p})/P_{\rm in}$ and reflection $R(\omega_{\rm p}) = P_{\rm R}(\omega_{\rm p})/P_{\rm in}$ spectra as functions of the frequency $\omega_p$ of the incident probe field $P_{\rm in}$. Probe spectra $\{T(\omega_{\rm p}),R(\omega_{\rm p})\}$ are recorded following the detection of a single-atom event with $\omega_{\rm p} = \omega_{\rm c}$ for a fixed detuning $\Delta_{\rm ca}$ between atom and cavity to optimize sensitivity for an intracavity atom (i.e., $\Delta_{\rm ca} = \Delta_{\rm pa}$). With an atom thereby localized in the cavity mode, fast control logic and feedback switch the probe power $P_{\rm in}$ to some fiducial level for a given spectrum and the probe frequency $\omega_{\rm p}$ to a relevant detuning $\Delta_{\rm pa} \neq \Delta_{\rm ca}$ for measurements of $\{T(\omega_{\rm p}),R(\omega_{\rm p})\}$. Because falling atoms remain in the evanescent field for only a few $\mu$s, the spectra are built up over thousands of transit detections and consequently represent an ensemble average over triggered atom trajectories.

\begin{figure}
\includegraphics[width=1.0\columnwidth]{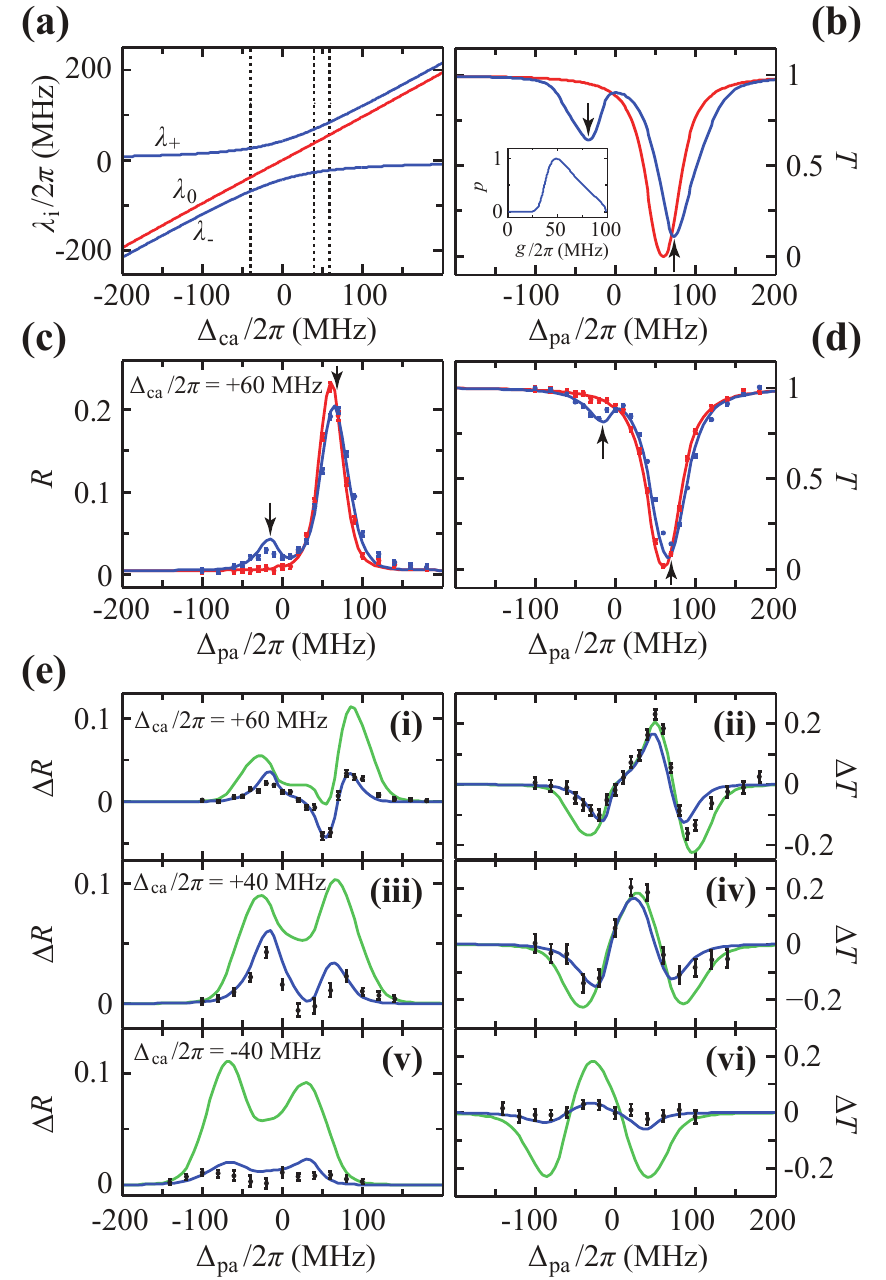}\vspace{-0.10in}\caption{\label{fig4}
\textbf{Transmission $T(\omega_{\rm p})$ and reflection $R(\omega_{\rm p})$ spectra for single atoms coupled to a microtoroidal resonator. a,} cQED eigenvalues $\lambda_{\pm,0}$ for $\{h, g\}/2\pi=\{10, 40\}$ MHz as a function of atom-cavity detuning $\Delta_{\rm ca}$. The dashed lines indicate the detunings for the spectra in the following panels. \textbf{b,} $T(\omega_{\rm p})$ for $\Delta_{\rm ca}/2\pi = +60$ MHz from a simple average for falling atoms over the distribution $p_{\rm fall}(g)$ (inset) absent cavity and surfaces forces.  $\Delta \omega_{\rm peaks}$ is found from the peaks indicated by arrows. \textbf{c, d,} Experimental reflection $R_i(\Delta_{\rm pa})$ and transmission $T_i(\Delta_{\rm pa})$ spectra for the empty cavity $i = {\rm NA}$ (red) and with atoms $i= {\rm A}$ (blue) with the peaks used for $\Delta \omega_{\rm exp }$ indicated. Curves are results of the full Monte Carlo simulation. \textbf{e}  Difference spectra $\Delta R = R_{\rm A}(\Delta_{\rm pa}) - R_{\rm NA}(\Delta_{\rm pa})$ and $\Delta T = T_{\rm A}(\Delta_{\rm pa}) - T_{\rm NA}(\Delta_{\rm pa})$ for $\Delta_{\rm ca}/2\pi = +60$ (i,ii), $+40$ (iii,iv), $-40$ MHz (v,vi). Green lines are simulation results for $U_{\rm s} =U_{\rm d}=0$, while blue lines are from the complete simulation.
}\end{figure}

Strong radiative coupling of an atom and a microtoroidal resonator is described by an extension of the Jaynes-Cummings \cite{JaynesCummings:1963} Hamiltonian~\cite{Aoki:2006} (see SI). Our whispering-gallery resonator supports two counter-propagating traveling-wave modes $\{a, b\}$ that are coupled by internal scattering at a rate $h$, leading to cavity eigenmodes $\{A,B\}$ that are linear superpositions of $\{a, b\}$. The interaction of the $\{A,B\}$ modes with an intracavity atom is characterized by coherent coupling $g(\vect{r})$, with the resulting atom-cavity eigenvalues $\{\lambda_i\}$ shown in Fig. \ref{fig4}a for the single-excitation eigenstates. For large detuning $|\Delta_{\rm ca}| \gg g$, there is one atom-like and two cavity-like (for the $\{A,B\}$ modes) eigenvalues. For $\Delta_{\rm ca} \sim g$, there is the familiar anti-crossing between the imaginary parts of two dressed-state eigenvalues $\lambda_\pm$ with splitting $\Delta \lambda_{\pm} = \text{Im}\left(\lambda_{+} - \lambda_{-}\right) \approx \sqrt{\Delta_{\rm ca}^2 + 4 g^2}$ for $g\gg \{h, \kappa, \gamma_{\|}\}$, while the third cavity-like eigenvalue $\lambda_0$ remains uncoupled to the atom. This dressed-state eigenstructure, along with the dissipative rates $\gamma_{\|}(d)$ and $\kappa$ for atom and cavity, determine the system's spectral response $\{T(\omega_{\rm p}),R(\omega_{\rm p})\}$.

Using a simple model with atoms falling vertically through the evanescent field of Fig. 1b with $\{U_{\rm s},U_{\rm d}\} = 0$ (SI), we construct a probability distribution $p_{\rm fall}(g)$ of coupling constants for atom detection, with probe spectra $\{T(\omega_{\rm p}),R(\omega_{\rm p})\}$ then obtained by averaging spectra for fixed $g$ over the distribution $p_{\rm fall}(g)$ (Fig. 4b). Although the full eigenstructure from Fig. 4a cannot be resolved due to the `smearing' from $p_{\rm fall}(g)$ even with $g_{\rm max} \gg \{\kappa,\gamma_{\|}\}$, the splitting $\Delta \omega_{\rm peaks}^{\rm (b)}$ between $\lambda_{-}$ and $\lambda_0$ (shifted by its proximity to the unresolved $\lambda_{+}$) is resolved and approximates, though underestimates, the eigenvalue splitting $\Delta \lambda_{\pm}$ (i.e., $\Delta \omega_{\rm peaks}^{\rm (b)}/2\pi = 110$ MHz, while $\Delta \lambda_{\pm}/2\pi = 130$ MHz).

%fig4

Figure \ref{fig4}c, d show measured spectra for both the bare-cavity with no atoms (NA), $R_{\rm NA}\left(\Delta_{\rm pa}\right)$ and $T_{\rm NA}\left(\Delta_{\rm pa}\right)$, and with triggered single atoms (A), $R_{\rm A}\left(\Delta_{\rm pa}\right)$ and $T_{\rm A}\left(\Delta_{\rm pa}\right)$, for $\Delta_{\rm ca}/2\pi = 60$ MHz. The splitting $\Delta \omega_{\rm exp}/2\pi \approx 95\pm 5$ MHz between the prominent cavity peak and the dressed state feature can be read directly from both $T_{\rm A}$ and $R_{\rm A}$. Taking $\Delta \omega_{\rm peaks}^{\rm ( c,d)}$ as a lower estimate for the average eigenvalue splitting $\Delta \lambda_{\pm}$ yields an average coupling $\overline{g}/2\pi \approx 37 \pm 3$ MHz.  This average coupling indicates that strong coupling is achieved on average, with $\overline{g} > \left(\kappa, \gamma_0\right)$, where $\left(\kappa, \gamma_0\right)/2\pi = \left(21, 2.6\right)$ MHz.

Quantitative differences between the simple model in Fig. \ref{fig4}b and the experimental spectra in Fig. \ref{fig4}c, d yield information about effects beyond the simple model, including perturbative surface interactions not included in the standard Jaynes-Cummings treatment\cite{Aoki:2006, Dayan:2008}. In particular, the feature at $\Delta_{\rm pa}/2\pi = -30$ MHz in both $R_{\rm A}$ and $T_{\rm A}$ is significantly reduced in magnitude from the spectrum predicted by $p_{\rm fall}(g)$, which as discussed below, results from the effects of $\{U_{\rm s},U_{\rm d}\}$ on the atomic trajectories and internal levels as in Figs. \ref{fig2}, \ref{fig3}.

Measurements of the difference spectra taken with and without atom transit events, $R_{\rm A}-R_{\rm NA}$ and $T_{\rm A}-T_{\rm NA}$ are shown in Figs. \ref{fig4}e for cavity detunings $\Delta_{\rm ca}/2\pi = 60, 40, -40$ MHz. Again, the simple prescription of reading $\Delta \omega_{\rm peaks}$ directly from the splitting of the low and high frequency peaks together with the expression for $\Delta \lambda_{\pm}$ leads to an estimate of the average coupling  $\overline{g}/2\pi \approx 35 \pm 5$  MHz that is consistent across the six spectra displayed.

For comparison to the measured spectra, the full curves in Figs. \ref{fig4}e are from our Monte Carlo simulation for $\Delta R \equiv R(\Delta_{\rm pa}) - R_{g\rightarrow 0}(\Delta_{\rm pa})$ and $\Delta T \equiv T(\Delta_{\rm pa}) - T_{g\rightarrow 0}(\Delta_{\rm pa})$. Calculated spectra are shown both for the full model and with all forces removed. Agreement with the full model is achieved for the choice $g_{\rm max}/2\pi \sim 100$ MHz, which is somewhat less than the value of $g_{\rm max}/2\pi =140$ MHz expected for the fundamental TE mode near $\lambda = 852$ nm estimated from a finite element calculation (Fig. \ref{fig1}b).  The difference is attributable to imprecise knowledge of the toroid geometry and mode. Except for the relevant detunings and measured cavity decay rates, the same parameters are used for each spectrum simulation; specifically, $\left(g_{\rm max}, \gamma_0\right)/2\pi = \left(100,  2.6\right)$ MHz. Note that apart from the adjustment of $g_{\rm max}$, all parameters in the simulation are estimated from measurements of our system or, in the case of dipole forces, surface forces, and surface level shifts, are taken from theoretical and experimental results in the literature (SI).

For each of the spectra in Fig.~\ref{fig4}e, removing the Casimir-Polder and dipole forces (i.e., setting $U_{\rm s},U_{\rm d}$ to zero) leads to increased deviations from the measured spectra relative to the full simulation, which describes the measurements reasonably well. The most significant effect of $U_{\rm s}$ is seen for a red-detuned cavity ($\Delta_{\rm ca}/2\pi = -40$ MHz) where significant spectral features not readily observed in the data appear for $U_{\rm s} =0$. In combination with the temporal analysis in Figs. \ref{fig2}, \ref{fig3}, the cQED spectra illustrate the necessity of including perturbative surface interactions for understanding atomic dynamics near the resonator. The model uses a distance dependent atomic decay rate $\gamma_{\|}(d)$ for the linearly polarized TE mode, but the differences between $\gamma_{\|}$ and $\gamma_{0}$ are too small to be observed in the data.  Despite the overall consistency achieved with the full simulation, systematic disagreements between data and model suggest that further analytical progress is required, including better (independent) knowledge of the toroid geometry, as well as solving the full master equation to account for the multi-level structure of the Cs atom~\cite{Birnbaum:2006}.

%%%%%%%%%%%%%%%%%%%%%%%%%%%%%%%%%%%%%%%%%%%%%%%%%%%%%%%%%%%%%%%%%%%%%%%%%%%%
%%%%%%%%%%%%%%%%%%%%%%%% Correlations Section %%%%%%%%%%%%%%%%%%%%%%%
%%%%%%%%%%%%%%%%%%%%%%%%%%%%%%%%%%%%%%%%%%%%%%%%%%%%%%%%%%%%%%%%%%%%%%%%%%%%

\begin{figure}[b]
\includegraphics[width=1.0\columnwidth]{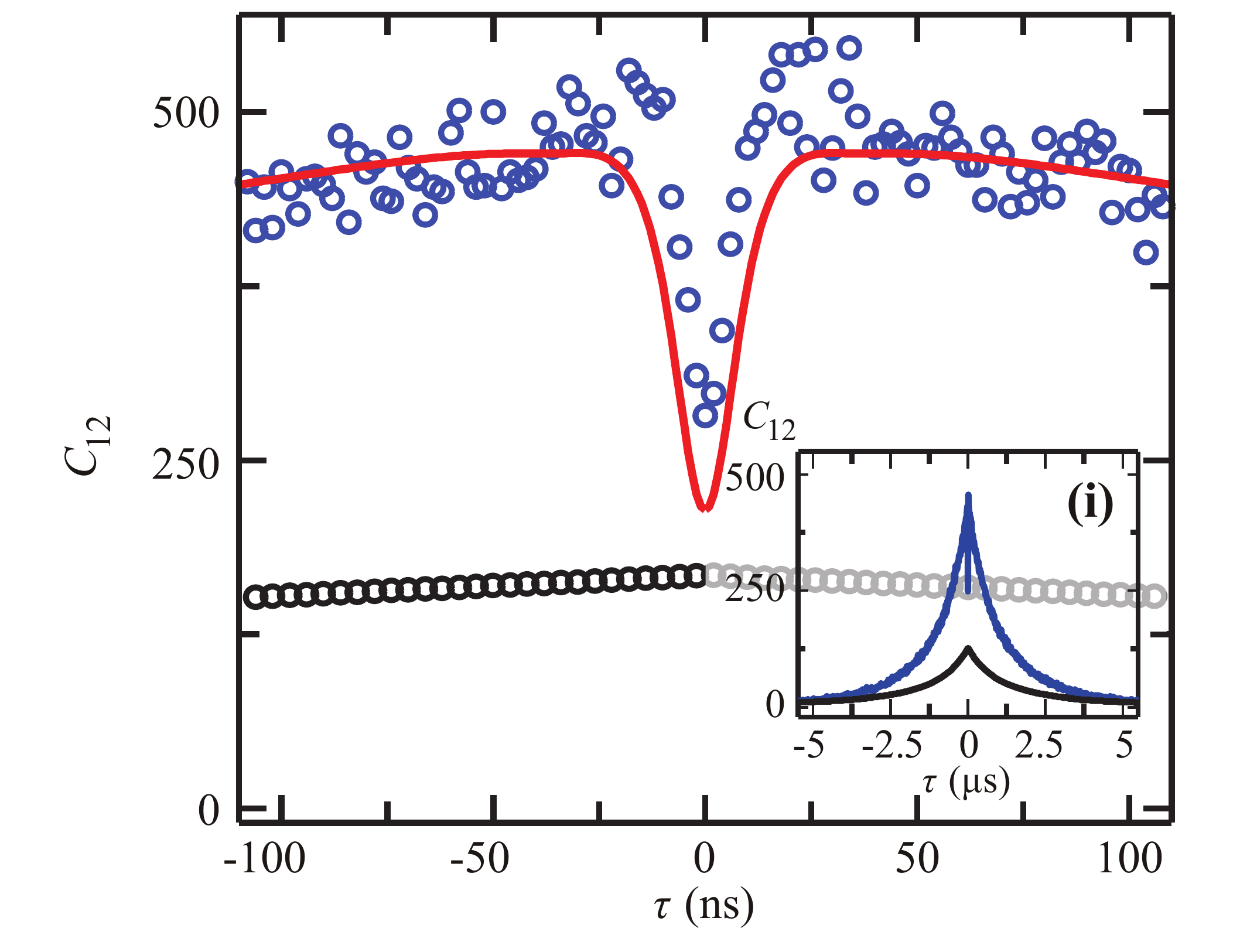}
\vspace{-0.25in}
\caption{\label{fig5}
\textbf{Photon statistics for atoms with $\Delta_{\rm ca} = 0, \Delta_{\rm pa} = 0$. }  Cross-correlation $C_{12}(\tau)$ (blue circles) computed from the records of photoelectric counts at detectors $D_1,D_2$ from the forward flux $P_{\rm T}$ from a sum over many atom trajectories showing photon antibunching around $\tau = 0$, with $\overline{C}_{12}(\tau)$ obtained from the product of averages of the recorded counts at each detector for comparison (black circles). The red curve is a calculation for the two-time second-order correlation function averaged over the distribution $p(g)$ from the full simulation and scaled by a single parameter to match $C_{12}(\tau)$ at $\tau = \pm40$ ns. (i)  Expanded view of $C_{12}(\tau)$ and $\overline{C}_{12}(\tau)$ over full range of $\tau$, with the long decay related to the typical atom transit time.}
\vspace{-0.1in}
\end{figure}

To confirm the quantum nature of the atom-cavity interaction near the surface of the toroid, we present in Fig. \ref{fig5} measurements of photon statistics for the transmitted field $P_{\rm T}$ for $\Delta_{\rm ca} = 0$. Photon statistics are inferred from the time records $C_{1,2}(t_i)$ of photoelectric counts in time bins $t_i$ for two detectors $D_1,D_2$ following an atomic trigger event at $t=0$.  Fig. \ref{fig5} shows the average cross-correlation $C_{12}(\tau) = \sum_i \langle C_{1}(t_i)C_{2}(t_i+\tau)\rangle$ for $0<t_i<8$ $\mu$s as well as the cross-correlation of the average counts $\overline{C}_{12}(\tau) = \sum_i \langle C_1(t_i)\rangle \langle C_2(t_i+\tau)\rangle$, where the angled brackets represent sums over the ensemble of triggers. The photon process is super-Poissonian, indicated by $C_{12}(\tau) > \overline{C}_{12}(\tau)$ for all $\tau$, due to large fluctuations in transmitted intensity from one atom to the next, which presumably arise from variations in atomic position $\vect{r}$ and hence coupling $g(\vect{r})$ near the toroid's surface (inset (i)). Even in the face of these large fluctuations, the non-classical character of the atom-cavity interaction survives, as is evident from the short-time dynamics shown in Fig. \ref{fig5}, where $C_{12}(0) < C_{12}(\tau)$ exhibits photon antibunching. From the minimum at $\tau = 0$, $C_{12}(\tau)$ regress to its peak with characteristic half-width of $6$ ns.

In contrast to the case $C_{12}\rightarrow 0$ realized in microtoroids in the bad-cavity limit for a photon turnstile \cite{Dayan:2008} and a photon router \cite{Aoki:2009}, here $C_{12}(0)$ is $0.55$ of its maximum value. This behavior is a result of coherent dressed-state dynamics for $\Delta_{\rm ca} = 0$ rather than a projective measurement of the atom as in Refs. \onlinecite{Dayan:2008,Aoki:2009}. Averaging a time-dependent calculation for the coincidence count rate for a fixed value of $g$ over the distribution of $g$ obtained from our trajectory simulations results in reasonable agreement with our measurements (red curve in Fig. \ref{fig5}). Except for the assumptions inherent in our simulation, the only free parameter in this calculation is the amplitude which is scaled to match the data.  Our model predicts both $C_{12}(0) \neq 0$ and the regression timescale near $\tau = 0$, which supports its effectiveness in describing the quantum behavior of the atom near the surface of the toroidal resonator.

%%%%%%%%%%%%%%%%%%%%%%%%%%%%%%%%%%%%%%%%%%%%%%%%%%%%%%%%%%%%%%%%%%%%%%%%%%%%
%%%%%%%%%%%%%%%%%%%%%%%% Trapping and Conclusions Section %%%%%%%%%%%%%%%%%%%%%%%
%%%%%%%%%%%%%%%%%%%%%%%%%%%%%%%%%%%%%%%%%%%%%%%%%%%%%%%%%%%%%%%%%%%%%%%%%%%%

By exploiting real-time triggering of single atoms, our experiment has realized a system where an atom's dynamics are governed by both its strong, single-photon interactions with the resonator's field and perturbative forces on classical atomic motion and internal level structure from proximity to the resonator's surface.  Entering this regime opens the door for quantitative study of dynamical Casimir-Polder forces in the strong-coupling limit \cite{Buhmann:2008, Scheel:2009}, which will require trapping atoms at short distance scales for extended interrogation times. In contrast to the standing-wave structure of a Fabry-Perot cavity\cite{Ye:2008}, mirotoroidal resonators offer the tantalizing possibility of radially confining an atom in a circular orbit around the toroid \cite{Mabuchi:1994, Vernooy:1997}, with initial results from our simulation shown in Fig.~\ref{fig3}c. In correspondence to the development of cQED to reach the regime of strong coupling with one trapped atom in a Fabry-Perot resonator \cite{Ye:2008}, the advances described here offer an important advance toward trapping and cooling of a single atom near the surfaces of micro- and nano-scopic optical resonators.

%%%%%%%%%%%%%%%%%%%%%%%%%%%%%%%%%%%%%%%%%%%%%%%%%%%%%%%%%%%%%%%%%%%%%%%%%%%%
%%%%%%%%%%%%%%%%%%%%%%%% Methods Section %%%%%%%%%%%%%%%%%%%%%%%
%%%%%%%%%%%%%%%%%%%%%%%%%%%%%%%%%%%%%%%%%%%%%%%%%%%%%%%%%%%%%%%%%%%%%%%%%%%%

%%%%%%%%%%%%%%%%%%%%%%%%%%%%%%%%%%%%%%%%%%%%%%%%%%%%%%%%%%%%%%%%%%%%%%%%%%%%
%%%%%%%%%%%%%%% Acknowledgments, Figures, and References Section %%%%%%%%%%%%%%
%%%%%%%%%%%%%%%%%%%%%%%%%%%%%%%%%%%%%%%%%%%%%%%%%%%%%%%%%%%%%%%%%%%%%%%%%%%%

\textbf{Acknowledgments}
We acknowledge financial support from NSF, DoD NSSEFF program, Northrop Grumman Aerospace Systems, ARO, and IARPA.  N.P.S. acknowledges support of the Caltech Tolman Postdoctoral Fellowship. H. L. thanks the Center for the Physics of Information. Toroid fabrication was done in the Kavli Nanoscience Institute.  The authors thank A. S. Parkins, J. Ye and P. Zoller for illuminating discussions.

\vspace{10 mm}

\textbf{Appendix}

Silica microtoroidal resonators with principal diameter $D_{\rm p}\simeq24$ $\mu$m and minor diameter $D_{\rm m} \simeq 3$ $\mu$m are fabricated using standard methods\cite{Armani:2003} and mounted in an ultra-high vacuum chamber with pressure below $10^{-9}$ torr.  The cavity resonance frequency $\omega_{\rm c}$ is tuned using the silicon substrate temperature to be near the $6S_{1/2}, F = 4\rightarrow 6P_{3/2}, F^{\prime} = 5$ transition of Cs at frequency $\omega^{(0)}_{\rm a}$.  Cesium atoms are cooled below 10 $\mu$K in a magneto-optical trap and transported by an optical conveyor belt $\sim 800$ $\mu$m above the toroid.\cite{Aoki:2009} The microtoroid is excited by a tapered fiber carrying an input power $P_{\rm in} \sim 4$ pW at frequency $\omega_{\rm p}$, which is coupled to a whispering-gallery mode of the toroid using piezoelectric positioners.  The cavity internal losses $\kappa_{\rm i}$, fiber-taper coupling $\kappa_{\rm ex}$, and intermodal coupling $h$ are $\{\kappa_{\rm i}, \kappa_{\rm ex},h\}/2\pi \approx \{8, 10, 13\}$ MHz, while the maximum atom-cavity coupling rate at the surface of the toroid is inferred to be $g_{\rm max}/2\pi \approx 100$ MHz (i.e., a maximum single-photon Rabi frequency $\Omega^{(1)}_{\rm max}/2\pi = 2 g_{\rm max}/2\pi  \approx 200$ MHz). The transmitted ($P_{\rm T} $) and reflected ($P_{\rm R} $) powers from the coupling fiber are each directed to two single photon counting modules (SPCM), $\{D_1,D_2\}$ and $\{D_{1r},D_{2r}\}$, respectively.  At critical coupling\cite{Spillane:2003}, the input field in the fiber taper and the field coupling out of the cavity destructively interfere in the forward transmission direction, so that $P_{\rm T} \lesssim 0.01 P_{\rm in}$ for $\omega_{\rm p} = \omega_{\rm c}$.  An atom coherently coupled to the cavity at rate $g(\vect{r})$ disturbs this critical coupling condition and the photodetector counts at $D_1,D_2$ increase. Single photoelectric events within a running time window $\Delta t_{\rm th}$ are counted and compared with a threshold number $C_{\rm th}$ by a field programmable gate array (FPGA) operating at $40$ MHz.  For the data shown in Figs. 3 and 4, the parameters $\Delta t_{\rm th} = 750$ ns and $C_{\rm th} = 5$ are chosen to give a false detection rate less than $1\%$ and an average trigger time as early as possible during the atom transits (SI). When the threshold criteria is met (defined as $t \equiv 0$ for each event), the FPGA sends a trigger pulse to a photon counting card to time stamp and record subsequent photodetections with 2-ns time resolution.

At $t = 0$, the FPGA also triggers high-frequency optical modulators that switch the power and frequency of the probe input to the tapered fiber.  For typical experiments, the probe flux is reduced to $P_{\rm in}\simeq 2$ pW and the probe detuning $\Delta_{\rm pa} = \omega_{\rm p} - \omega^{(0)}_{\rm a}$ set to a value within the range shown in Fig. \ref{fig4}. Including both optical and electrical delays, the optical switching is complete by $t = 100$ ns.  The photocount record on detector $i$, $C_i(t)$, in a time interval $0< t < 8$ $\mu$s following the trigger is recorded for a succession of $N \gg 1$ trigger events.  For spectral measurements, the transmitted and reflected photocount records are averaged over a selected time window (typically, $200 < t < 700$ ns) and normalized by the photocounts taken with large detuning, $P_{\rm T}\left(\Delta_{\rm pa} \gg \kappa \right)$, to obtain the experimental transmission and reflection spectra, $T\left(\Delta_{\rm pa}\right)$ and $R\left(\Delta_{\rm pa}\right)$. Error bars for data are estimated assuming Poissonian counting statistics and are written as plus or minus one s.d.  Fit results are quoted with 68\% confidence intervals.
%For further details, refer to SI.

Having validated our trajectory simulation with the measurements in Figs. \ref{fig2}-\ref{fig5}, we have studied loading of falling atoms into a two-color evanescent field far off-resonance trap (FORT) (Fig.~\ref{fig3}c) \cite{Mabuchi:1994, Vernooy:1997,Rosenblit:2006, Vetsch:2010}.  A trapping potential $U_{\rm t}$ can be formed using a blue-detuned fundamental mode and a higher order red-detuned mode \cite{Mabuchi:1994, Vernooy:1997}.  For our simulation, we use a red (blue)-detuned mode near 898 nm (848 nm) with powers $\sim 50$ $\mu$W each to give a trap depth of $\sim 1.5$ mK which is switched on at $t=0$ conditioned on a falling atom FPGA trigger.  Despite the large kinetic energy of falling atoms and poor localization of the atoms relative to the trap minimum, approximately $20\%$ of triggered atom trajectories are captured in the trap. Simulated trapping times exceed 50 $\mu$s, limited not by heating from trapping light but by the radiation pressure from unbalanced traveling whispering-gallery modes.  This radiation pressure leads to atom gallery orbits around the toroid (Fig.~\ref{fig3}c).  Exciting a red-detuned standing wave would provide three-dimensional trap confinement and increase the trap lifetime.


\begin{thebibliography}{99}

\bibitem{Miller:2005} Miller,~R., \textit{et al}. Trapped atoms in cavity QED: coupling quantized light and matter.  \textit{J. Phys. B: At. Mol. Opt. Phys.} \textbf{38}, S551-S565 (2005).

\bibitem{Kimble:2008} Kimble, H.~J. The quantum internet. \textit{Nature} \textbf{453}, 1023-1030 (2008).

\bibitem{Aoki:2006}  Aoki, T., \textit{et al}.  Observation of strong coupling between one atom and a monolithic microresonator. \textit{Nature} \textbf{442}, 671-674 (2006).

\bibitem{HarocheKleppner:1989} Haroche, S. and Kleppner, D. Cavity Quantum Electrodynamics. \textit{Physics Today}, \textbf{42} 24-30 (1989).

\bibitem{Casimir-Polder:1948} Casimir, H. B. G. and Polder, D. The influence of retardation on the London-van der Waals forces. \textit{Phys. Rev.}, \textbf{73}, 360-372 (1948).

\bibitem{Purcell:1946} Purcell, E. M. Spontaneous emission probabilities at radio frequencies.  \textit{Phys.~Rev.~} \textbf{69}, 681 (1946).

\bibitem{Drexhage:1974}  Drexhage, K. H. Interaction of light with monomolecular dye layers.  \textit{Progress in Optics}, volume XII, ed. E. Wolf, 163-232. Elsevier, New York (1974).

\bibitem{Gabrielse:1985}  Gabrielse, G and Dehmelt, H.  Observation of inhibited spontaneous emission.   \textit{Phys.~Rev.~Lett.} \textbf{55}, 67-70 (1985)

\bibitem{Hulet:1985}  Hulet, R.G., Hilfer, E. S., and Kleppner, D.  Inhibited spontaneous emission by a Rydberg atom.   \textit{Phys.~Rev.~Lett.} \textbf{55},  2137-2140 (1985)
\bibitem{Sukenik:1993} Sukenik, C. I., Boshier, M. G., Cho, D., Sandoghar, V., and Hinds, E. A.  Measurement of the Casimir-Polder force.  \textit{Phys. Rev. Lett.} \textbf{70}, 560 (1993).

\bibitem{Berman:1994} Berman, P. \textit{Cavity Quantum Electrodynamics,} (San Diego: Academic Press, 1994).

\bibitem{Odom:2006} Odom, B., Hanneke, D., D'Urso, B., and Gabrielse, G. New measurement of the electron magnetic moment using a one-electron quantum cyclotron. \textit{Phys.~Rev.~Lett.} \textbf{97}, 030801 (2006).

\bibitem{devices}  Michler, P., \textit{et al}.  A quantum dot single-photon turnstile device.  \textit{Science} \textbf{290}, 2282-2285 (2000).

\bibitem{JaynesCummings:1963}  Jaynes, E.T., and Cummings, F. W. Comparison of quantum and semiclassical radiation theories with application to the beam maser, \textit{Proc . IEEE}, \textbf{51}, 89 (1963).

\bibitem{Meschede:1985} Meschede,~D., Walther, H., and M\"{u}ller, G. One-atom maser. \textit{Phys.~Rev.~Lett.}~\textbf{54}, 551-554 (1985).

\bibitem{Thompson:1992} Thompson, R. J., Rempe, G., and Kimble, H. J. Observation of normal-mode splitting for an atom in an optical cavity. \textit{Phys. Rev. Lett.}~\textbf{68}, 1132-1135 (1992).

\bibitem{HarocheBook}  Haroche, S. and Raimond, J.-M. \textit{Exploring the Quantum: Atoms Cavities, and Photons.}  (Oxford: Oxford Univeristy Press, 2006).

\bibitem{Vahala:2004} For a review, see Vahala, K. J.  Optical microcavities. \textit{Nature} \textbf{424}, 839-846 (2004).

\bibitem{Khitrova:2006} For a review, see Khitrova, G., Gibbs, H.M., Kira, M., Koch,~S.~W., and Scherer,~A. Vacuum Rabi splitting in semiconductors. \textit{Nature Physics} \textbf{2}, 81-90 (2006).

\bibitem{Schoelkopf:2008} For a review, see Schoelkopf, R. J. and Girvin, S. M. Wiring up quantum systems. \textit{Nature} \textbf{451}, 664-669 (2008).

\bibitem{Hofheinz:2008} Hofheinz, M., \textit{et al}. Generation of Fock states in a superconducting quantum circuit.  \textit{Nature} \textbf{434}, 310-314 (2008).

\bibitem{McKeever:2003} McKeever, J., Boca, A., Boozer, A. D., Buck, J. R., and Kimble,~H.~J.  Experimental realization of a one-atom laser in the regime of strong coupling.  \textit{Nature} \textbf{425}, 268-271 (2003).

\bibitem{Rempe:2009}  Weber, B., \textit{et al}.  Photon-photon entanglement with a single trapped atom. \textit{Phys. Rev. Lett.} \textbf{102}, 030501 (2009).

\bibitem{DiCarlo:2009}  DiCarlo, L., \textit{et al}.  Demonstration of two-qubit algorithms with a superconducting quantum processor.  \textit{Nature} \textbf{460}, 240-244 (2009).


\bibitem{Folman:2002} Folman, R., Kr\"{u}ger, P., Schmiedmayer, J., Denschlag, J., and Henkel, C. Microscopic atom optics: from wires to an atom chip.  \textit{Adv. At., Mol., Opt. Phys.} \textbf{48}, 263-356 (2002).

\bibitem{Armani:2003}  Armani, D. K., Kippenberg, T. J., Spillane, S. M., and Vahala, K. J.  Ultra-high-Q toroid microcavity on a chip. \textit{Nature} \textbf{421}, 925-928 (2003).

\bibitem{Spillane:2005} Spillane, S. M., \textit{et al.}  Ultrahigh-Q toroidal microresonators for cavity quantum electrodynamics. \textit{Phys. Rev. A} \textbf{71}, 013817 (2005).

\bibitem{Eichenfeld:2009} Eichenfield, M., Camacho, R., Chan, J., Vahala, K. J., and Painter, O.  A pictogram- and nanometer-scale photonic-crystal optomechanical cavity. \textit{Nature} \textbf{459}, 550-556 (2009).

\bibitem{SimPaper}  N.P. Stern, D. J. Alton, and H.J. Kimble, in preparation (2010).

\bibitem{Dayan:2008} Dayan, B., \textit{et al}.  A photon turnstile dynamically regulated by one atom. \textit{Science} \textbf{319}, 1062-1065 (2008).

\bibitem{Aoki:2009}  Aoki, T., \textit{et al}.  Efficient routing of single photons by one atom and a microtoroidal cavity. \textit{Phys. Rev. Lett.} \textbf{102}, 083601 (2009).

\bibitem{Mabuchi:1994}  Mabuchi, H. and Kimble, H.J.   Atom galleries for whispering atoms: binding atoms in stable orbits around an optical resonator.  \textit{Opt. Lett.} \textbf{19}, 749-751 (1994).

\bibitem{Vernooy:1997}  Vernooy, D. W. and Kimble, H. J. Quantum structure and dynamics for atom galleries.  \textit{Phys. Rev. A} \textbf{55}, 1239 (1997).

\bibitem{Rosenblit:2006}  Rosenblit, M., Japha, Y., Horak, P., and Folman, R.  Simultaneous optical trapping and detection of atoms by microdisk resonators.  \textit{Phys. Rev. A} \textbf{73}, 063805 (2006).

\bibitem{Birnbaum:2006} Birnbaum, K. M., Parkins, A. S., and Kimble, H. J. Cavity QED with multiple hyperfine levels. \textit{Phys. Rev. A} \textbf{74}, 063802 (2006).

\bibitem{Buhmann:2008} Buhmann, S. Y. Casimir-Polder forces on excited atoms in the strong atom-field coupling regime. \textit{Phys. Rev. A} \textbf{77}, 012110 (2008).

\bibitem{Scheel:2009}  Scheel, S. and Buhmann, S. Y. Casimir-Polder forces on moving atoms. \textit{Phys. Rev. A} \textbf{80}, 042902 (2009).

\bibitem{Ye:2008} Ye, J. Kimble, H. J., and Katori, H. Quantum state engineering and precision metrology using state-insensitive light traps. Science \textbf{320}, 1734 (2008).


\bibitem{Spillane:2003} Spillane, S. M., Kippenberg, T.J., Painter, O. and Vahala, K. J. Ideality in a fiber-taper-coupled microresonator system for application to cavity quantum electrodynamics. \textit{Phys. Rev. Lett}. \textbf{91}, 043902 (2003).


\bibitem{Vetsch:2010} Vetsch, E., Reitz, D., Sagu\'{e}, Schmidt, R., Dawkins, S. T., and Rauschenbeutel, A.  Optical interface created by laser-cooled atoms trapped in the evanescent field surrounding an optical nanofiber.  \textit{Phys. Rev. Lett.} \textbf{104}, 203603 (2010).


\end{thebibliography}
\end{document}